\documentclass[a4paper,11pt,fleqn]{article}

\usepackage{amsmath}					%Additional math environments
\usepackage{amssymb}					%Additional math symbols
\usepackage{caption}
\usepackage{graphicx}
\usepackage{subfig}
\usepackage[singlespacing]{setspace}
\usepackage{titlesec}  					%Change spacing before and after headings
\usepackage{enumitem}					%for customizing iteminze and enumerate
\usepackage[english]{babel}		%hyphenation support
\usepackage{siunitx}					%Display units correctly with space between number and unit
\usepackage[]{prettyref}				%comfortable referencing of tables, figures and equations
\usepackage{authblk}

\addto\captionsenglish{}	%Change caption for figures
\titlespacing{\section}{0pt}{6pt}{0pt}			%includes medskipamount of 6pt --> 12pt space above, 6pt space below
\titlespacing{\subsection}{0pt}{0pt}{0pt}		%includes medskipamount of 6pt --> 6pt space above, 6pt space below
\titlespacing{\subsubsection}{0pt}{0pt}{0pt}	%includes medskipamount of 6pt --> 6pt space above, 6pt space below
\titleformat{\section}{\normalfont\fontsize{14}{18}\bfseries}{\thesection}{1em}{}		%change font size for section
\titleformat{\subsection}{\normalfont\fontsize{12}{14}\bfseries}{\thesection}{1em}{}		%change font size for subsection
\titleformat{\subsubsection}{\normalfont\fontsize{11}{13}\bfseries}{\thesection}{1em}{}		%change font size for subsubsection

\newcommand*{\TitleFont} {\usefont{\encodingdefault}{\rmdefault}{b}{n}\fontsize{14}{18} \selectfont}	%use T1 as encoding parameter, otherwise no bold font
\newcommand*{\EPEtitle}[1]{\title{\TitleFont #1}}			%define title

\newcommand*{\EPEkeyword}[1]{$\ll$#1$\gg$}					%define macro for keywords

\newenvironment{EPEreference}{\begin{thebibliography}{99} \setlength{\itemsep}{-6pt}  \fontsize{10}{12}\mdseries } {\end{thebibliography}}

\newrefformat{fig}{Fig.~\ref{#1}}
\newrefformat{tab}{Table~\ref{#1}}

\usepackage{latexrelease}

\EPEtitle{Loss Minimization of Traction Systems in Battery Electric Vehicles Using Variable DC-link Voltage Technique --- Experimental Study\vspace{0.2cm}}
\vspace{0.75cm}
\author[$\dagger$]{Libo Liu}
\author[$\dagger$]{Boyang Li}
\author[$\dagger$$\dagger$]{Gunther G\"otting}
\author[$\dagger$$\dagger$$\dagger$]{Yusheng Xiang}
\author[$\dagger$]{Qusay Salem}
\author[$\dagger$]{Muhammad Hamid}
\author[$\dagger$]{Jian Xie}
\affil[$\dagger$]{Institute for Energy Conversion and Storage, Ulm University, 89081 Ulm, Germany, Phone: +49 (0)731 5025542, Email: libo.liu@uni-ulm.de}
\affil[$\dagger$$\dagger$]{Powertrain Solutions, Control, Robert Bosch GmbH, 70442 Stuttgart, Germany, Email: Gunther.Goetting@de.bosch.com}
\affil[$\dagger$$\dagger$$\dagger$]{Institute of Vehicle System Technology,
	Karlsruhe Institute of Technology, 76131 Karlsruhe, Germany, Email: yusheng.xiang@partner.kit.edu}

\date{}										%print no date

\begin{document}
	\maketitle	
	\section*{Acknowledgments}
	This work is funded by Robert Bosch GmbH of Germany under the contract of No. B-111225C.
	\section*{Keywords}
	%Please use \EPEkeyword to include your keywords to ensure correct format
	\EPEkeyword{Permanent magnet motor}, \EPEkeyword{Multilevel converters}, \EPEkeyword{Efficiency}, \EPEkeyword{Electrical drive}, \EPEkeyword{Electric vehicle}, \EPEkeyword{Battery Management Systems (BMS)}. 
	\section*{Abstract}

A novel variable dc-link voltage technique is proposed to reduce the traction losses for electrical drive applications. A 100-unit cascaded multilevel converter is developed to generate the variable dc-link voltage. Experimental measurement shows that the machine additional losses and IGBT-inverter losses are reduced substantially. The system efficiency enhancement is at least 2\%.

\section*{Introduction}
\begin{figure}[!t]
	\centering 
	\includegraphics[scale=0.82]{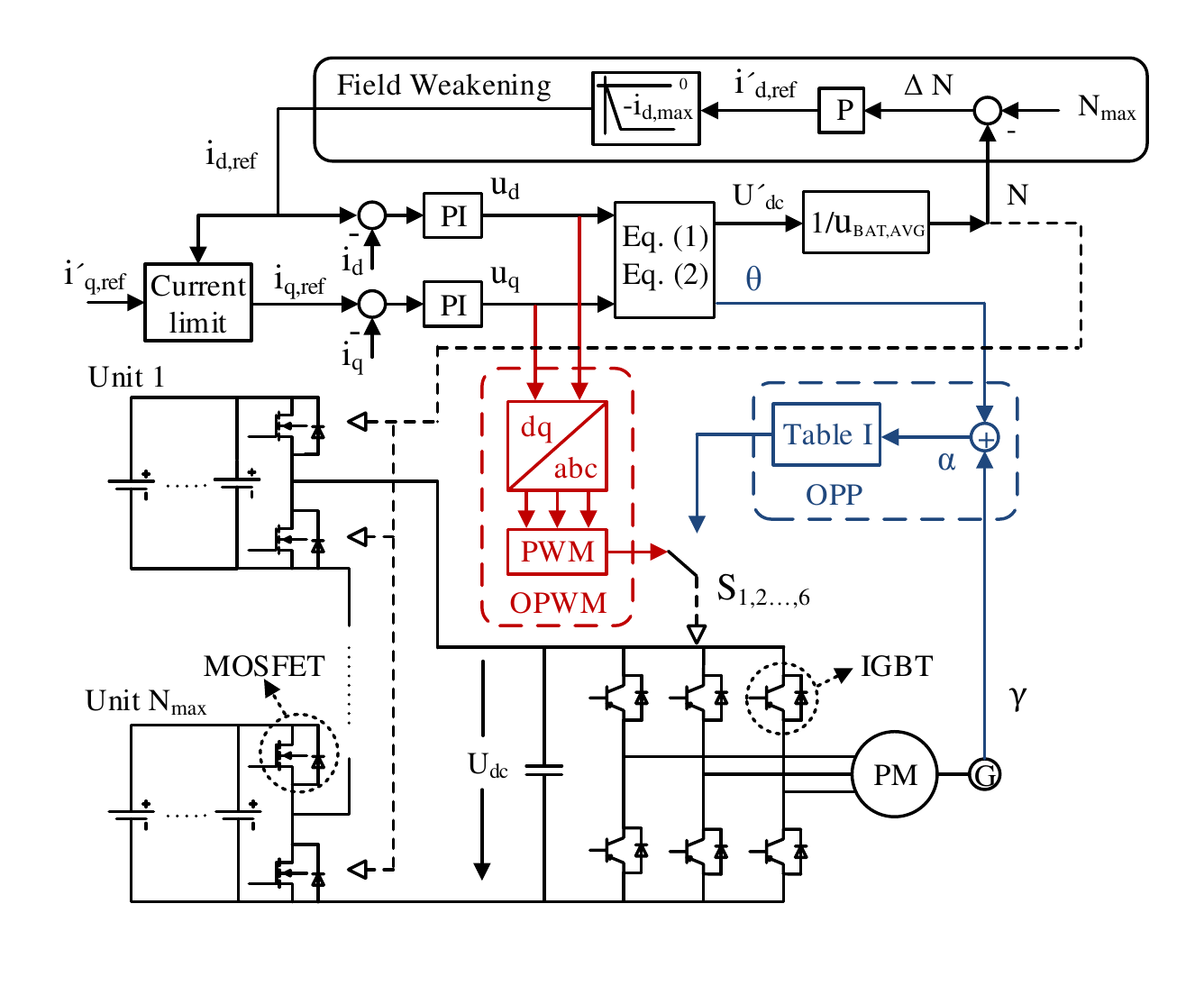}
	\caption{Configuration and control scheme of the cascaded multilevel converter, using the variable dc-link voltage technique}
	\label{fig:CMI_Control}
\end{figure}

\begin{figure}[!b]
	\centering 
	\includegraphics[scale=0.6]{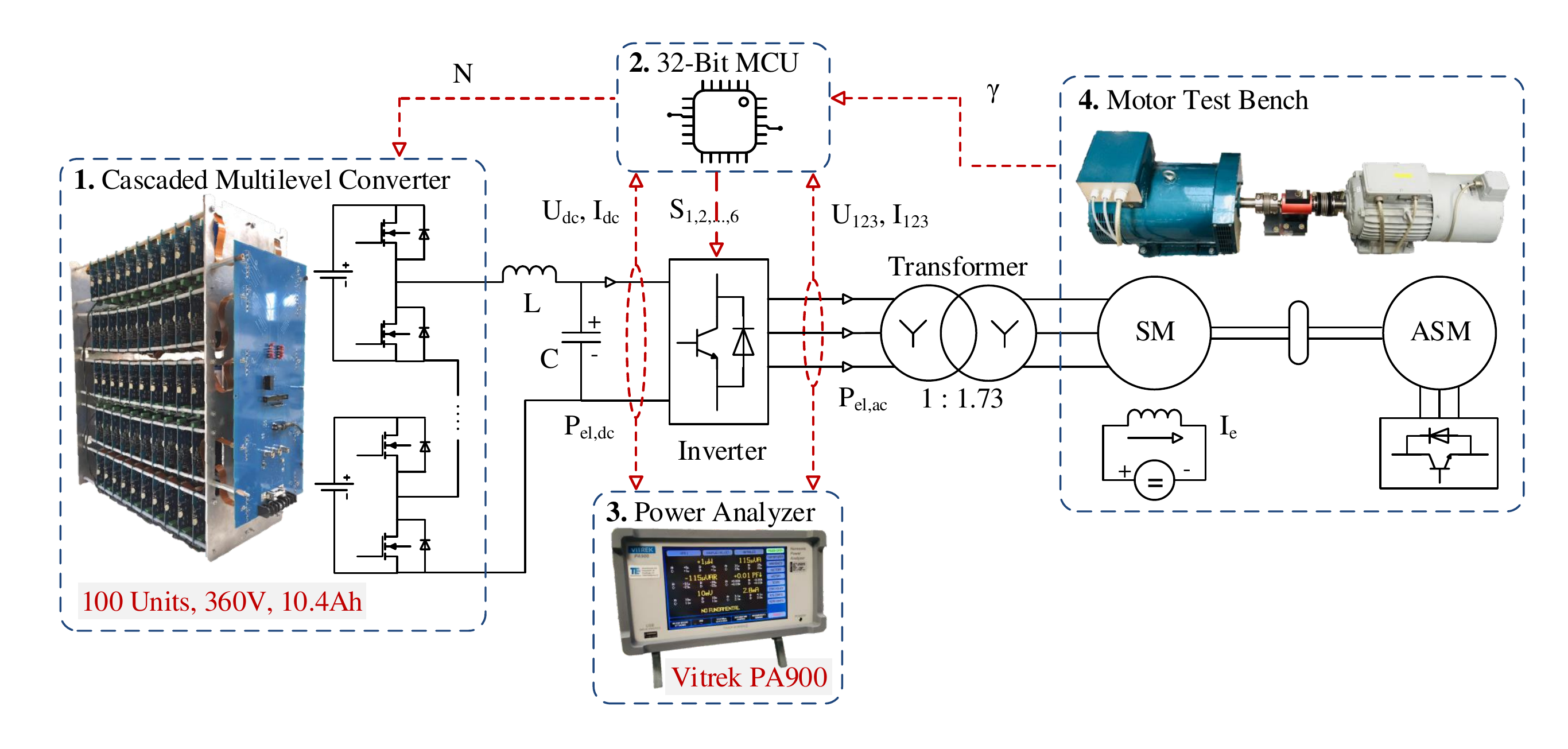}
	\caption{Experimental setup for measurements of traction losses}
	\label{fig:exp_setup}
\end{figure}

In general, the battery pack of an electric vehicle (EV) consists of multiple cells that are direct in series and parallel connection \cite{Miao_2019}. The major drawback of this topology is that the dc-link voltage remains unchanged over the full speed/torque range. The maxed-out dc-link voltage results in large machine and inverter losses, particularly at low speeds \cite{Liu_2018}. It is investigated that the carrier harmonics of the PWM inverter contribute a significant part to the core losses of an electrical machine \cite{Heseding_2016}\cite{Yamazaki_2009}. To minimize the overall traction losses, a variable dc-link voltage, which can adapt to different speed/torque conditions, is preferable.\\
This can be realized by connecting a bidirectional dc/dc converter between the battery pack and traction inverter. It has been clarified in \cite{Yu_2013} that the desired dc-link voltage at low speeds is much lower than that for high-speed operation. By using the dc/dc converter, the energy saving is substantial, mainly due to the decrease of switching losses and core losses \cite{Tenner_2012}\cite{Najmabadi_2015}. However, at high speeds, the dc/dc converter has shown disadvantages of efficiency because of the losses caused by the additional converter \cite{Najmabadi_2015}. Besides, the bulky passive components of the converter must be taken into account for vehicle applications that have restricted space envelope. Although the arguments for having the bidirectional dc/dc converter, it is clear that the adjustable dc-link voltage can improve the efficiency of the drive train. If the additional losses introduced by the converter can be minimized, a high-efficiency drive over the full speed range can be achieved.\\ 
Inspired by the application of dc/dc converter for the battery pack, this work utilizes the cascaded multilevel converter (CMC) to produce the variable dc-link voltage. Compared with the conventional dc/dc converter, the CMC has higher efficiency due to the reduced stress of switching devices and the absence of the bulky passive components (inductors and capacitors). Moreover, the CMC is able to perform active balancing among the battery cells. Our previous calculation \cite{Liu_2018} has justified the potential of CMC for power loss reduction of the traction system. High-efficiency optimal modulation methods have been suggested for electric drives. In this work, a CMC prototype, consisting of 100 modular units, is developed to validate the proposed approach in \cite{Liu_2018}. The schematic diagram of the proposed CMC together with the traction inverter is shown in Fig.~\ref{fig:CMI_Control}. The CMC prototype can be seen on the left part of Fig.~\ref{fig:exp_setup}. Each modular unit contains a half-bridge circuit and two NMC battery cells in parallel connection. The MOSFETs in the half-bridge circuit allow the cells to be connected to the dc-link or bypassed. In this manner, the dc-link voltage can be controlled. In the meanwhile, charge equalization among these cells can be realized by arranging the charge/discharge period of the individuals. Because there are always idle cells if the dc-link voltage doesn't max out.\\
In order to evaluate the impact of the proposed variable dc-link voltage technique on the traction losses, the measurement results using the optimal modulation methods are compared with the results using the conventional PWM method (constant dc-link voltage). It is verified that the variable dc-link technique has effectively reduced the machine losses and inverter losses. The efficiency enhancement of 2\% is guaranteed over the full speed range.
\section*{Modulation methods and motor control scheme}
The proposed modulation methods for the variable dc-link voltage are the optimal PWM (OPWM) and optimal pulse pattern (OPP). The former is essentially the conventional PWM method combined with an adjustable dc-link voltage $U_{dc}$. This enables the modulation index $m_a$ to approach 1 for low-speed operation. The optimal pulse pattern is also called selective harmonic elimination (SHE). However, the conventional application of SHE, where a constant dc-link voltage is used, requires multiple switching patterns to achieve a continuous output of fundamental voltage \cite{Yang_2016}. By contrast, the OPP scheme is able to control the dc-link voltage via the CMC according to the desired operating point. To avoid confusion with the SHE used in other applications, the SHE used here, featuring a variable dc-link voltage, is called OPP. Compared to the PWM method, the OPP has a lower inverter switching frequency, which is proportional to the fundamental frequency. Table~\ref{tab:9-chop} lists the switching angles of a 9-pulse OPP within $\pi/2$ period, by which the harmonics up to 29th-order can be eliminated.\\
The motor control scheme is based on the field oriented control, as depicted in Fig.~\ref{fig:CMI_Control}. $i_{d,ref}$ and $i_{q,ref}$ are the flux and torque generating components of the stator current. $u_{d}$ and $u_{q}$ denote the \textit{d-q} stator voltages, which are generated from the current controllers. Then, the dc-link voltage $U'_{dc}$ is obtained using \eqref{eq:u1_angle} and \eqref{eq:udc_cal},\\
\begin{equation}\label{eq:u1_angle}
U_1=\sqrt{u_d^2 + u_q^2},
\ \ \theta =\arctan(u_q/u_d),\\
\end{equation} 
\begin{equation}\label{eq:udc_cal}
U'_{dc}=2\cdot U_1/u_{r},
\end{equation} 
where $U_1$ represents the fundamental amplitude of the stator voltage $\underline{U}_1$, $\theta$ is the electrical angle of $\underline{U}_1$ with respect to $d$-axis. $u_{r}$ denotes the dc-link voltage utilization ratio. Depending on which modulation method is used, $u_{r}$ can be different. For the OPWM method, $u_{r}$ equals the modulation index $m_a$, which is set to 0.9 in this work. Within the scheme of the OPP, $u_{r}$ equals the generalized fundamental value $u_{1,gen}$ of the switching pattern. The optimal switching angles and the corresponding $u_{1,gen}$ can be calculated according to \cite{Chiasson_2003}. Typically, the OPP features a higher voltage utilization ratio than the OPWM, because $u_{1,gen}$ is usually greater than 1. Once $U'_{dc}$ is available, the required number of battery cells $N$ can be obtained by dividing the average battery terminal voltage. For any desired stator voltage $\underline{U}_1$, the angle of $\underline{U}_1$ in the stationary frame, $\alpha$, is equal to the sum of $\theta$ and the electrical angle of $d$-axis $\gamma$. Afterwards, $\alpha$ determines the switching states of the three-phase inverter following the OPP.\\
For normal operation, $i_{d,ref}$ is set to zero to maximize the torque per ampere ratio \cite{Nalepa_2012}. As the motor accelerates, the back-EMF gets larger and a high dc-link voltage is required. When $N$ increases to the maximum number $N_{max}$, field weakening is enabled by producing a negative $i_{d,ref}$. It should be noted that cell balancing is only performed in non-field-weakening area. As long as $N$ is smaller than $N_{max}$, the battery cells with low capacity can be put in idle status. 
\begin{table}[t]
	\centering
	\caption{Generalized fundamental value and switching angle in $\mathrm{rad}$ for 9-pulse OPP}
	\label{tab:9-chop}
	\begin{tabular}{ |l|l|l|l|l|l|l|l|l|l| } 
		\hline
		\textbf{$U_{1,gen}$} &\textbf{$\alpha_1$} & \textbf{$\alpha_2$} & \textbf{$\alpha_3$} & \textbf{$\alpha_4$} & \textbf{$\alpha_5$}&\textbf{$\alpha_6$} & \textbf{$\alpha_7$} & \textbf{$\alpha_8$} &\textbf{$\alpha_9$}\\ 
		\hline
		1.1597 & 0.0811 & 0.1882 & 0.2409 & 0.3862 & 0.4212&0.5761 & 0.5946 & 1.3219 & 1.3282\\ 
		\hline
	\end{tabular}
\end{table}
\section*{Experimental setup}
\begin{figure}[!b]
	\subfloat[DC-link voltage and rotation speed\label{subfig:udc_wm_vs_t}]{%
		\includegraphics[width=0.48\textwidth]{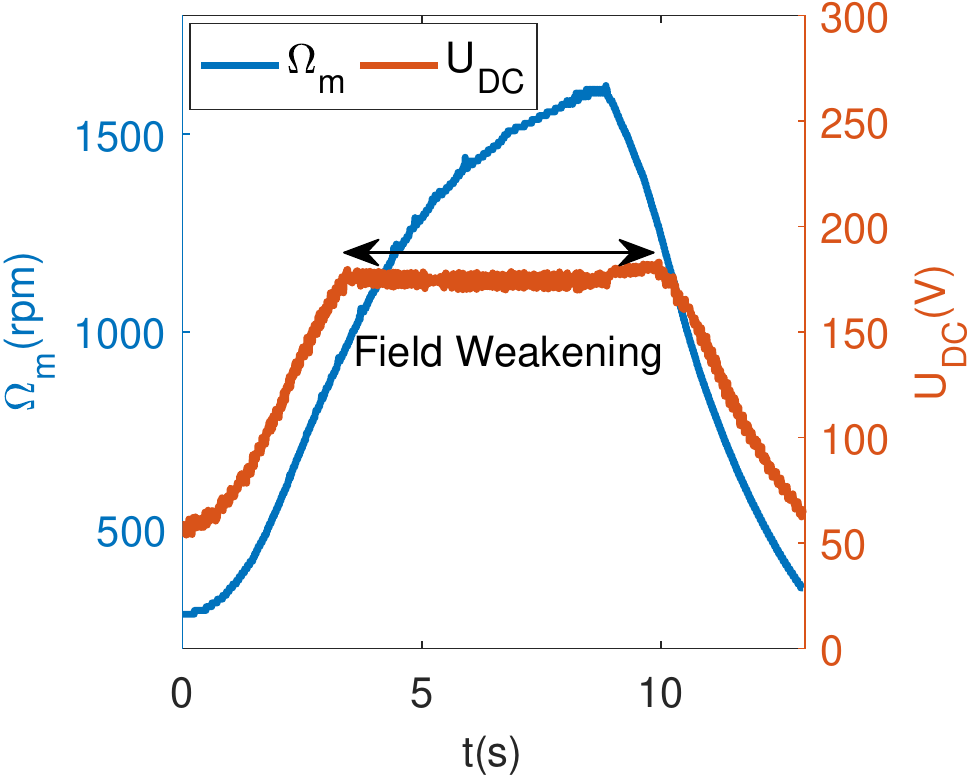}
	}
	\hfill
	\subfloat[$dq$-component of stator current\label{subfig:idq_vs_t}]{%
		\includegraphics[width=0.48\textwidth]{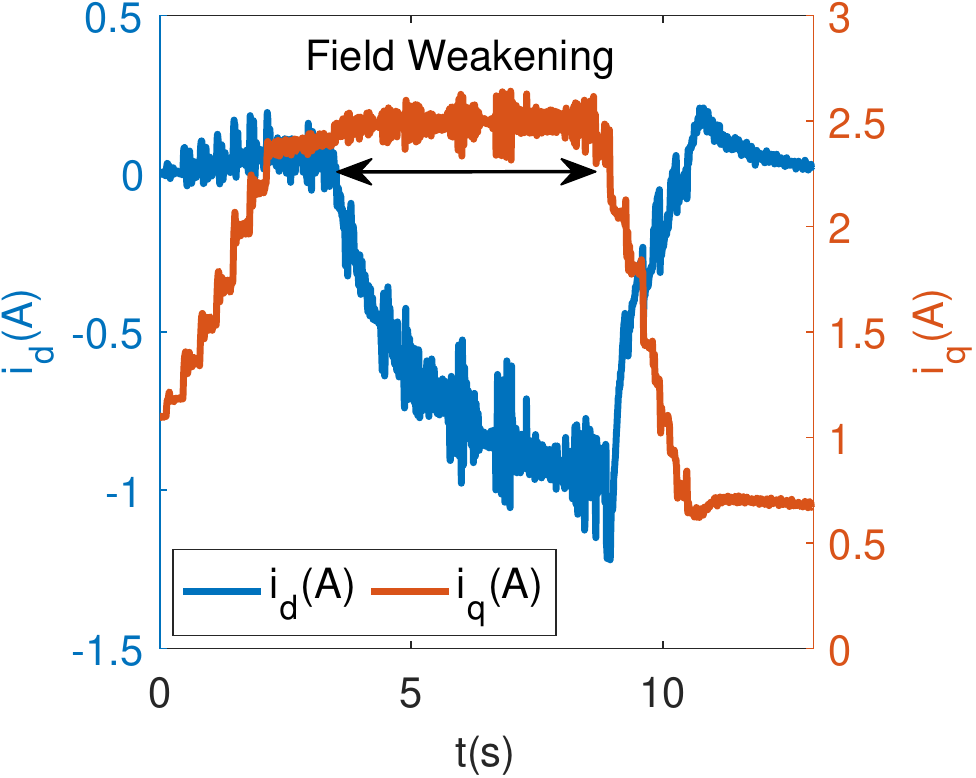}
	}
	\caption{A drive scenario using the variable dc-link voltage technique OPWM, $N_{max} = 50$}
	\label{fig:var_udc_t}
\end{figure}

\begin{figure}[!htpb]
	\subfloat[$U_{dc}$ versus $\Omega_m$\label{subfig:udc_vs_wm}]{%
		\includegraphics[width=0.48\textwidth]{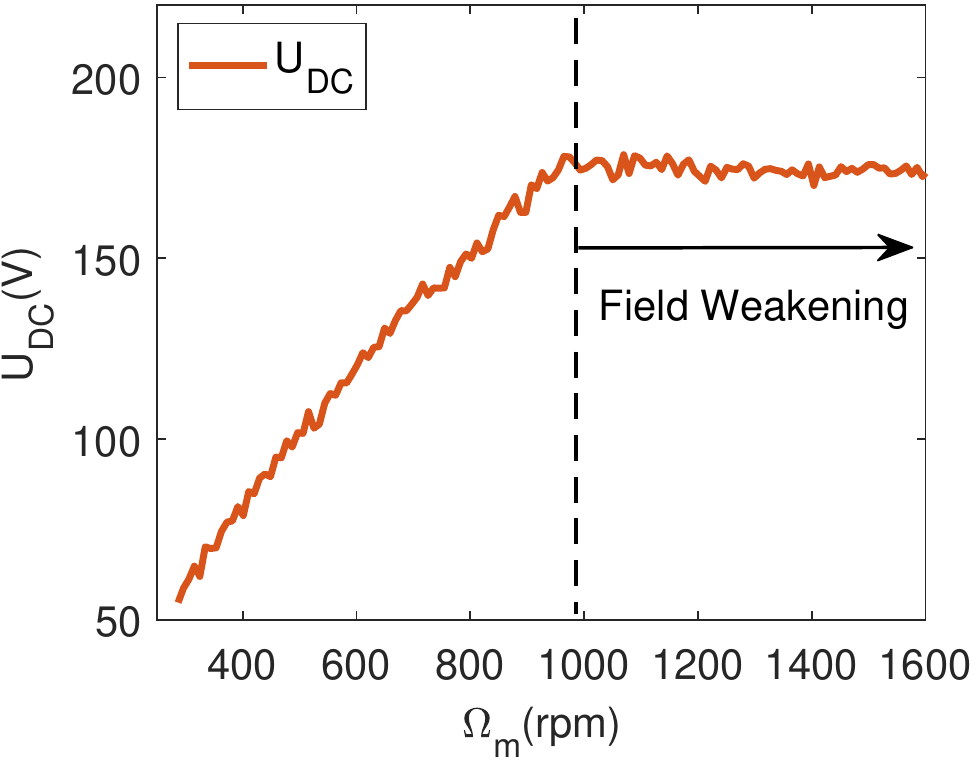}
	}
	\hfill
	\subfloat[$i_{d}$, $i_{d}$ versus $\Omega_m$\label{subfig:idq_vs_wm}]{%
		\includegraphics[width=0.48\textwidth]{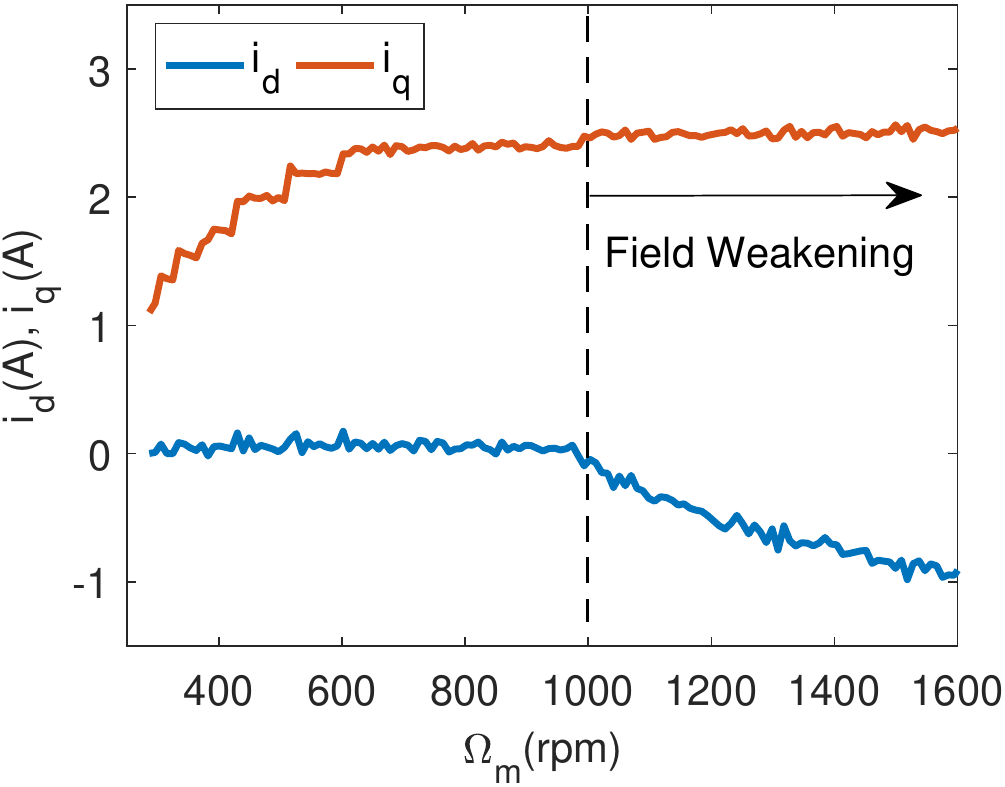}
	}
	\caption{DC-link voltage and $dq$-current over the full speed range}
	\label{fig:var_udc_wm}
\end{figure}

Fig.~\ref{fig:exp_setup} shows the experimental setup to evaluate the impact of different modulation methods on traction losses. The cascaded multilevel converter consists of 100 modular switchable units. Each unit contains 2 parallel connected NMC batteries. The cell used here is TR-26650LI that has a capacity of 5200mAh, 3.6V nominal voltage and 3C/16.6A maximal current output. Thus, the total capacity of the CMC is 360V, 10.4Ah. The power analyzer \textit{Vitrek-PA900} has 4 power measurement channel cards providing 0.1\% basic accuracy with 1Mhz class bandwidth. The test motor is a 4-pole synchronous motor \textit{Rotek STC-5} with the ratings of 400V, 5kW, 9A and 1500rpm. The asynchronous machine \textit{Lenze DFRARS132-22} acts as a load.\\
A drive scenario using the OPWM method is demonstrated in Fig.~\ref{fig:var_udc_t} and \ref{fig:var_udc_wm}. The motor control algorithm is described in Fig.~\ref{fig:CMI_Control}. The motor test bench is a bit obsolete. For safety purposes, overspeed (above 1500$rpm$) should be avoided when running the machine. In order to enter the field-weakening mode in advance (above 1000$rpm$), $N_{max}$ is set to 50. Results in Fig.~\ref{fig:var_udc_t} and \ref{fig:var_udc_wm} have justified the feasibility of the proposed variable dc-link voltage technique. For normal operation, the dc-link voltage increases with the speed. When the motor operates in the field weakening mode, the dc-link voltage maxes out and keeps constant.

\section*{Categories of traction losses}
Besides the two proposed optimal modulation methods, as a comparison, the conventional PWM method with the constant dc-link voltage is also investigated to evaluate the traction losses. The traction losses discussed in this work consist of the machine additional losses, inverter losses, MOSFET losses and battery losses. 
\subsection*{Machine additional losses}
Machine additional losses are caused by the harmonics of the inverter output voltage and current. They are affected by the dc-link voltage and inverter switch pattern \cite{Yamazaki_2009}. The power flow of an electrical machine is described as
\begin{equation}\label{eq:p_flow}
P_{el,ac} = P_{mot,f1} + P_{mot,fh} + P_{mech} + P_{friction} + P_{windage},
\end{equation} 
where $P_{mot,f1}$ represents the machine losses due to the fundamental waveform \{$U_1$, $I_1$\} and is the sum of iron losses and copper loss. $P_{mot,fh}$ denotes the machine additional losses, which involves the additional iron losses due to voltage harmonics and additional copper loss due to current harmonics. $P_{mech}$ is the mechanical power, which is the product of the rotation speed and mechanical torque. $P_{friction}$ and $P_{windage}$ are the friction loss and windage loss. The former is proportional to the rotational speed and the latter is proportional to its cube. In practice, it is difficult to measure $P_{mot,fh}$ directly, therefore, a comparative measurement is made. When the motor is adjusted at the same operating point, i.e., the same \{$\Omega_m$, $M$, $U_1$, $I_1$\}, these power losses ($P_{mot,f1}$, $P_{mech}$, $P_{friction}$ and $P_{windage}$) are considered unchanged for all three modulation schemes. Consequently, the variation of the inverter output power $P_{el,ac}$, that can be read from the power analyzer, implies an increase or a decrease in the machine additional losses.
\subsection*{Inverter losses}
The IGBT adopted here is \textit{Infineon FS50R06KE3} IGBT Module. The inverter losses ($P_{inv}$) comprise the switching and conduction losses of the IGBTs. $P_{inv}$ can be obtained by subtracting the inverter output power $P_{el,ac}$ from the input power $P_{el,dc}$. They can be read from the power analyzer.

\subsection*{MOSFET losses}
The MOSFET losses consist of the switching and conduction losses. Because the modular unit in the cascaded converter is either bypassed (lower MOSFET on) or connected (upper MOSFET on). There are always 100 MOSFETs in conduction. The MOSFET used here is \textit{PSMN1R0-30YLC} with the on-state resistance of 0.85m$\Omega$. Switching losses occur when N varies. The control frequency of the dc-link voltage is set to 5kHz. During the control period, N is allowed to increase or decrease by 1. Thus, the maximum MOSFET switching frequency is 5kHz. The switching voltage equals the terminal voltage of a single battery cell.
\subsection*{Battery losses}
The battery losses are due to the internal resistance that can be divided into DC and AC components. Since the battery current for steady-state is considered as constant, only DC part is taken into account. The DC resistance of a single cell used here is 40m$\Omega$. The total battery losses of the CMC is calculated as follows,
\begin{equation}\label{eq:p_bat}
P_{bat} = N\cdot I_{dc}^2\cdot 40m\Omega/N_p,
\end{equation} 
where $N_p$ is the parallel number of the cells in each modular units. $N_p$ equals 2 in the prototype.

\section*{Evaluation of Traction losses}
\begin{table}[!t]
	\centering
	\caption{Operating points (OPs) for power loss evaluation, $M$ = 11.5Nm, $N_{max}$ = 90}
	\label{tab:op_data}
	\begin{tabular}{ |l|l|l|l|l|l|l| } 
		\hline
		& OP1 & OP2 & OP3 & OP4 & OP5 & OP6 \\ 
		\hline
		$U_1^{rms}$ & 73.8V & 93.2V & 109.1V & 120.0V & 130.2V & 138.9V \\ 
		\hline
		$I_1^{rms}$ & 6.4A & 6.5A & 6.8A & 7.2A & 7.8A &  8.3A \\ 
		\hline
		cos$\phi$ & 0.9703 & 0.9877 & 0.9945 & 0.9981 & 0.9986 & 0.9969 \\ 
		\hline
		$\Omega_m$ & 549 rpm & 759 rpm & 948 rpm & 1134 rpm & 1332 rpm & 1503 rpm \\ 
		\hline	
	\end{tabular}
\end{table}
\begin{table}[!t]
	\centering
	\caption{Measured electric parameters using 3 modulation methods for 6 OPs}
	\label{tab:meas_data}
	\begin{tabular}{ |l|l|l|l|l|l|l|l|l|l| } 
		\hline
		& \multicolumn{3}{c|}{$U_{dc}$} & \multicolumn{3}{c|}{$P_{el,dc}$} & \multicolumn{3}{c|}{$P_{el,ac}$}\\
		\cline{2-10}
		& PWM & OPWM & OPP & PWM & OPWM & OPP & PWM & OPWM & OPP \\ 
		\hline
		OP1 & 325V & 153V & 108V  & 876 W & 846 W & 840 W & 825 W & 810 W & 812 W\\ 
		\hline
		OP2 & 319V & 191V & 135V  & 1143 W & 1104 W & 1096 W & 1090 W & 1063 W & 1066 W\\ 
		\hline
		OP3 & 314V & 223V & 157V  & 1374 W & 1346 W & 1329 W & 1320 W & 1300 W & 1297 W\\ 
		\hline
		OP4 & 309V & 248V & 173V & 1628 W & 1603 W & 1575 W & 1570 W & 1550 W & 1540 W \\ 
		\hline
		OP5 & 303V & 265V & 189V & 1891 W & 1859 W & 1831 W & 1830 W & 1800 W & 1790 W\\ 
		\hline
		OP6 & 297V & 273V & 202V & 2115 W & 2083 W & 2063 W & 2050 W & 2020 W & 2013 W\\ 
		\hline		
	\end{tabular}
\end{table}
\begin{figure}[!b]
	\subfloat[$\Delta \eta$ due to reduction of $P_{mot,fh}$ \label{subfig:delta_motor}]{%
		\includegraphics[width=0.47\textwidth]{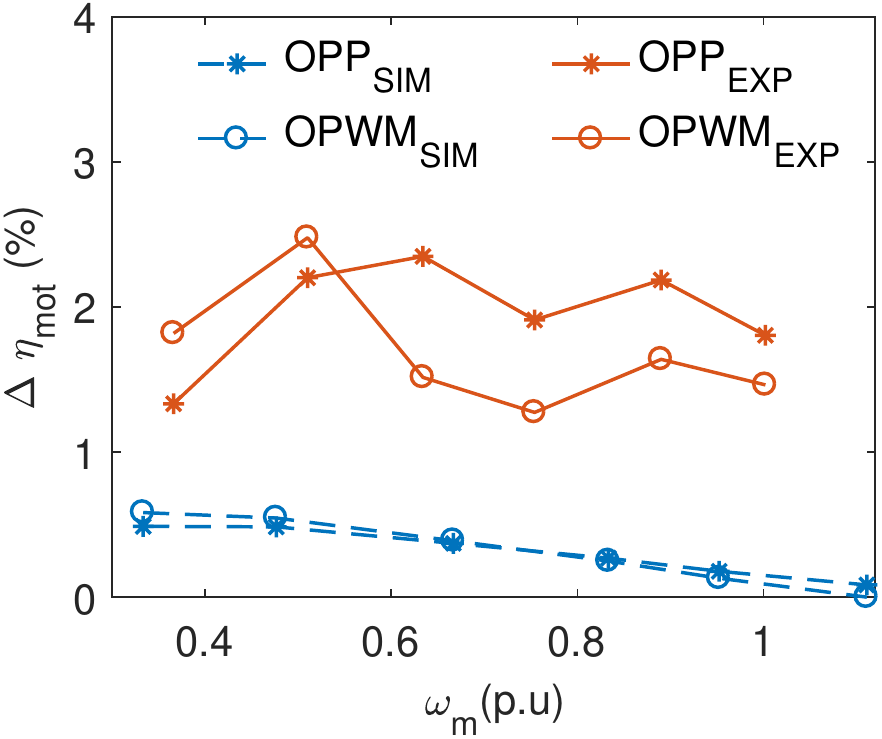}
	}
	\hfill
	\subfloat[$\Delta \eta$ due to reduction of $P_{mot,fh}$ and $P_{inv}$ \label{subfig:delta_motor_inv}]{%
		\includegraphics[width=0.47\textwidth]{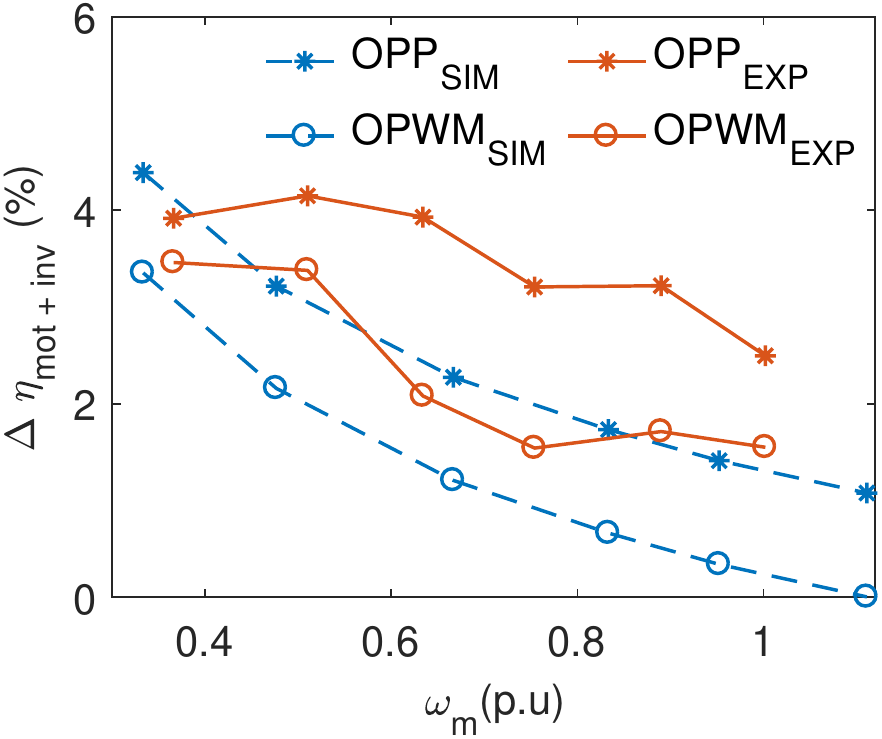}
	}
	\caption{Efficiency enhancement $\Delta \eta$ by using the variable dc-link voltage technique, compared to the conventional PWM method --- $SIM$: simulation results, $EXP$: experimental results}
	\label{fig:eff_imprv}
\end{figure}
To investigate the impact of modulation methods on the traction losses, 6 operating points (OPs) are applied to the test motor. Table~\ref{tab:op_data} lists the electrical parameters of the OPs. On each OP, measurement is made by using the conventional PWM, the OPWM, and the OPP method, respectively. $N_{max}$ is set to 90 for all measurements and $N$ is maintained at 90 in the case of PWM method.  In addition, the IGBT switching frequency of both PWM and OPWM methods is 10kHz. By contrast, the OPP shown in Table~\ref{tab:9-chop} has a lower switching frequency, which is 19 times the fundamental frequency. The measurement results are shown in Table~\ref{tab:meas_data}. The efficiency enhancement of the traction system by using the variable dc-link voltage technique is shown in Fig.~\ref{fig:eff_imprv}. The results match the simulation well. The simulation is done in \cite{Liu_2018}. Bear in mind that, the efficiency enhancement $\Delta\eta$ is with respect to the conventional PWM scheme using the constant dc-link voltage. The following formula is used to calculate $\Delta\eta$ for a given OP
\begin{equation}\label{eq:delta_eta}
\Delta\eta=\Delta P / P_{el,dc}^{pwm},
\end{equation} 
where $\Delta P$ represents the decrease of power losses due to the OPWM or OPP method. $P_{el,dc}^{pwm}$ is the total power consumption for the case of PWM scheme.\\
Fig.~\ref{subfig:delta_motor} shows that the system efficiency has increased by 2\% due to the reduction of machine additional losses. This can also be observed from the measured inverter output power $P_{el,ac}$ in Table~\ref{tab:meas_data}. For a given OP, the higher $P_{el,ac}$ implies more machine additional losses. The reason is that both OPWM and OPP methods require lower dc-link voltage in comparison to the conventional PWM. Therefore, the amplitude of the inverter voltage harmonics is reduced. Fig.~\ref{subfig:delta_motor_inv} demonstrates that the efficiency is further improved when taking the reduction of inverter losses into account --- 4\% at low speeds and 2\% at high speeds. Besides, the OPP method performs better than the OPWM from this perspective. Because the OPP features a higher voltage utilization ratio and a lower IGBT switching frequency, resulting in lower inverter losses.
\begin{figure}[!t]
	\subfloat[$\Delta \eta$ considering MOSFETs losses\label{subfig:delta_motor_inv_mft}]{%
		\includegraphics[width=0.47\textwidth]{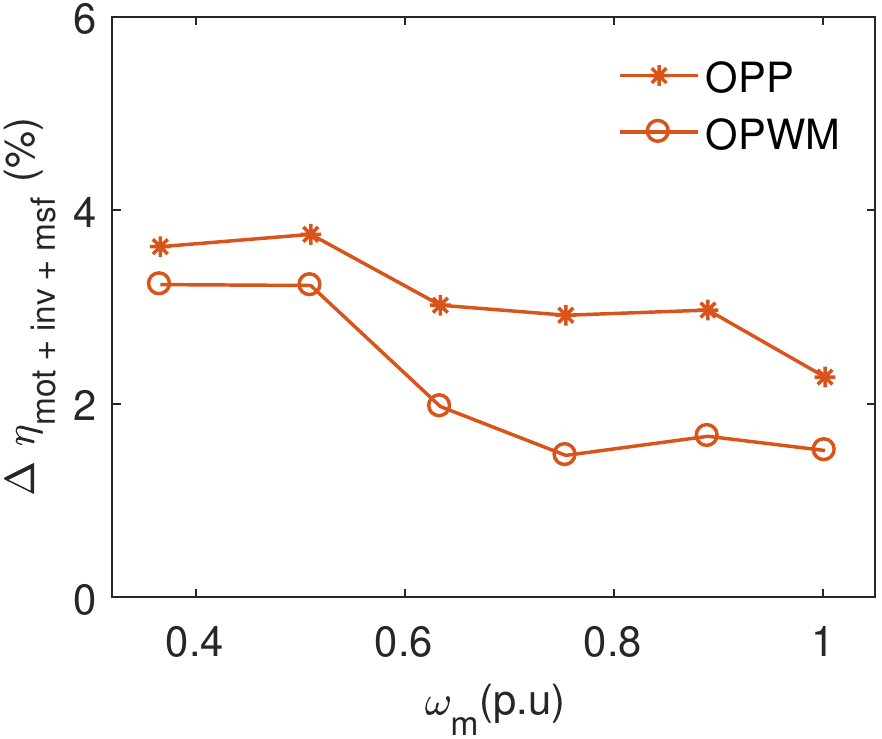}
	}
	\hfill
	\subfloat[$\Delta \eta$ considering MOSFETs and battery losses\label{subfig:delta_total}]{%
		\includegraphics[width=0.47\textwidth]{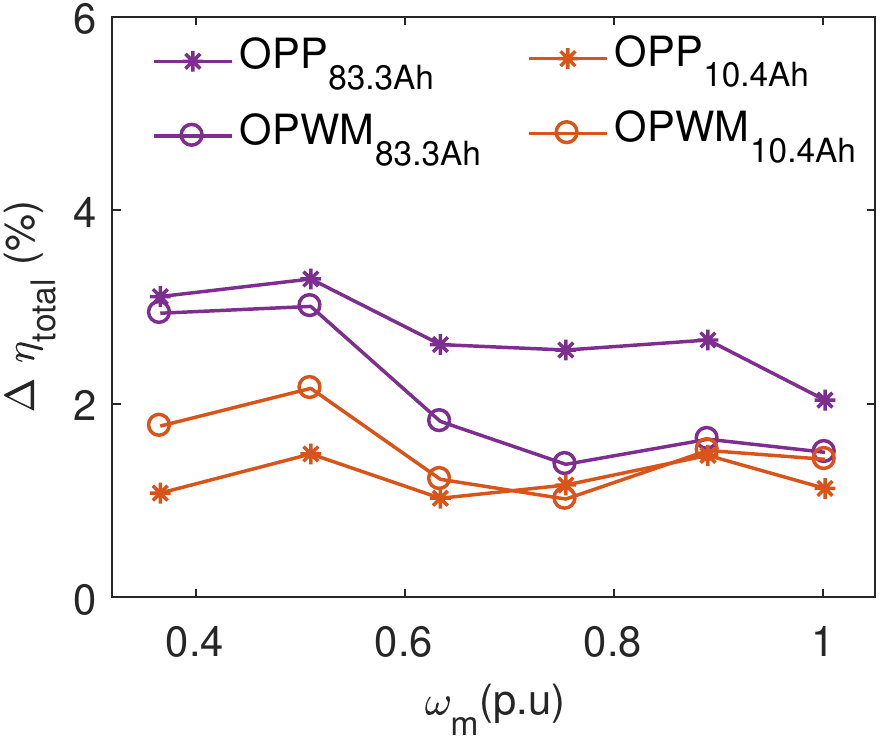}
	}
	\caption{Efficiency enhancement $\Delta \eta$ taking MOSFETs and battery losses into account}
	\label{fig:eff_imprv_total}
\end{figure}

\begin{figure}[!b]
	\centering 
	\includegraphics[scale=0.75]{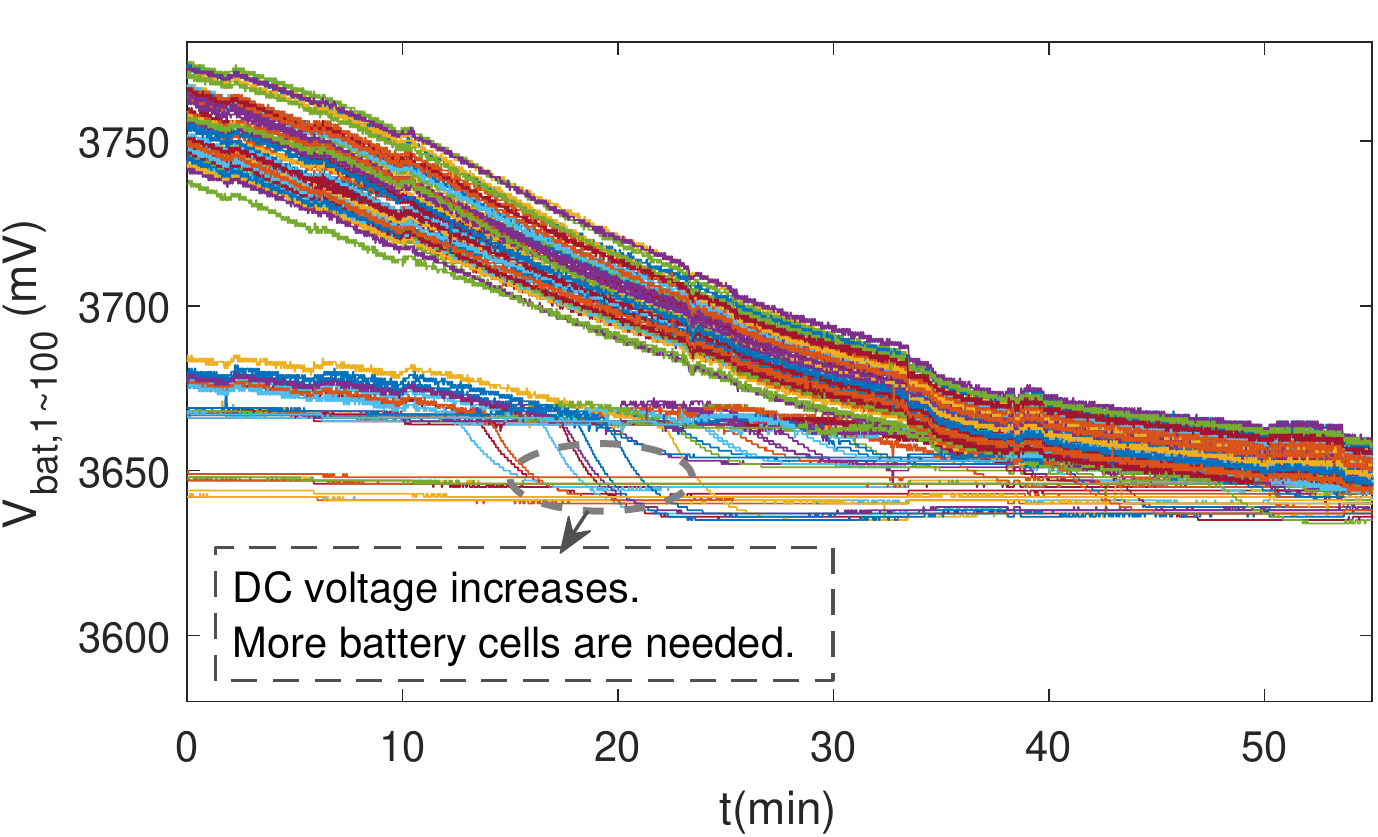}
	\caption{Balancing performance of the CMC with 4.2A discharge current, motor mode operation}
	\label{fig:dc_balancing}
\end{figure}
MOSFETs and battery losses are also evaluated, as shown in Fig.~\ref{fig:eff_imprv_total}. Since it is difficult to measure them directly, calculation results are used. The results of Fig.~\ref{subfig:delta_motor_inv_mft} indicate that the losses caused by the MOSFETs are very small compared with other losses. The reason is that the MOSFET has a low on-state resistance (0.85m$\Omega$) and switching voltage (3.6V). Fig.~\ref{subfig:delta_total} shows that the efficiency enhancement has dropped to 1\% when considering the battery losses. The compromise is due to the increase of the dc-current $I_{dc}$ for the same power output, hence more battery losses. In this work, the capacity of the battery pack is 360V, 10.4Ah. The equivalent DC resistance of the battery cells in a modular unit is 20m$\Omega$ . In practice, the capacity of the battery pack in an electric vehicle is higher than that of the prototype. Assuming a practical battery pack with 360V, 83.3Ah capacity, the equivalent resistance of the battery cells will be very small because there are multiple cells in parallel connection. From Fig.~\ref{subfig:delta_total}, it is clarified that the system efficiency is improved by at least 2\% when using batteries with high capacity. It can be concluded that high capacity is preferable to reduce battery losses. Fig.~\ref{fig:dc_balancing} shows the balancing performance of the CMC. The strategy is to let the batteries with low capacity relax when $N$ is smaller than $N_{max}$. In this scenario, the initial variance of the battery terminal voltage is 130mV. After 50min, the variance converges to 30mV.

\section*{Conclusion}

In this work, a variable dc-link voltage technique is presented to reduce the traction losses for electrical drive applications. Two optimal modulation methods are developed to regulate the dc-link voltage for various driving conditions. The experimental results have shown the machine additional losses and inverter losses can be substantially reduced in comparison to the conventional PWM method. The system efficiency increases by 4\% in the low-speed region and 2\% at high speeds. When taking battery losses into account, the efficiency enhancement is around 2\%. Furthermore, the developed cascaded multilevel inverter shows satisfactory performances on charge equalization among the battery cells. Compared to the bidirectional dc/dc topology, the proposed CMC is able to retain high efficiency for high-speed conditions. Due to the nature of modularization, an auxiliary battery management system for the conventional powertrain technology can be omitted. 

	\begin{EPEreference}

\bibitem{Miao_2019}
Y. Miao, P. Hynan, A. von Jouanne, A. Yokochi. ``Current Li-ion battery technologies in electric vehicles and opportunities for advancements," Energies, vol. 12, no. 6, p. 1074, January 2019.\vspace{3pt}

\bibitem{Liu_2018}
L. Liu, G. G\"otting and J. Xie, ``Loss Minimization Using Variable DC-Link Voltage Technique for Permanent Magnet Synchronous Motor Traction System in Battery Electric Vehicle," 2018 IEEE Vehicle Power and Propulsion Conference (VPPC), Chicago, IL, 2018, pp. 1-5. \vspace{3pt}

\bibitem{Heseding_2016}
J. Heseding, F. Mueller-Deile and A. Mertens, ``Estimation of losses in permanent magnet synchronous machines caused by inverter voltage harmonics in driving cycle operation," 2016 18th European Conference on Power Electronics and Applications (EPE'16 ECCE Europe), Karlsruhe, 2016, pp. 1-9. \vspace{3pt}

\bibitem{Yamazaki_2009}
K. Yamazaki and A. Abe, ``Loss Investigation of Interior Permanent-Magnet Motors Considering Carrier Harmonics and Magnet Eddy Currents," in IEEE Transactions on Industry Applications, vol. 45, no. 2, pp. 659-665, March-april 2009. \vspace{3pt}

\bibitem{Yu_2013}
C. Yu, J. Tamura and R. D. Lorenz, ``Control method for calculating optimum DC bus voltage to improve drive system efficiency in variable DC bus drive systems," 2012 IEEE Energy Conversion Congress and Exposition (ECCE), Raleigh, NC, 2012, pp. 2992-2999. \vspace{3pt}

\bibitem{Tenner_2012}
S. Tenner, S. Gimther and W. Hofmann, ``Loss minimization of electric drive systems using a DC/DC converter and an optimized battery voltage in automotive applications," 2011 IEEE Vehicle Power and Propulsion Conference, Chicago, IL, 2011, pp. 1-7. \vspace{3pt}

\bibitem{Najmabadi_2015}
A. Najmabadi, K. Humphries and B. Boulet, ``Implementation of a bidirectional DC-DC in electric powertrains for drive cycles used by medium duty delivery trucks," 2015 IEEE Energy Conversion Congress and Exposition (ECCE), Montreal, QC, 2015, pp. 1338-1345. \vspace{3pt}

\bibitem{Yang_2016}
K. Yang, Q. Zhang, J. Zhang, R. Yuan, Q. Guan, W. Yu and J. Wang, ``Unified Selective Harmonic Elimination for Multilevel Converters," in IEEE Transactions on Power Electronics, vol. 32, no. 2, pp. 1579-1590, Feb. 2017. \vspace{3pt}
	
\bibitem{Chiasson_2003}
J. Chiasson, L. M. Tolbert, K. McKenzie and Z. Du, ``A complete solution to the harmonic elimination problem," Eighteenth Annual IEEE Applied Power Electronics Conference and Exposition, 2003. APEC '03., Miami Beach, FL, USA, 2003, pp. 596-602 vol.1. \vspace{3pt}

\bibitem{Nalepa_2012}
R. Nalepa and T. Orlowska-Kowalska, ``Optimum Trajectory Control of the Current Vector of a Nonsalient-Pole PMSM in the Field-Weakening Region," in IEEE Transactions on Industrial Electronics, vol. 59, no. 7, pp. 2867-2876, July 2012. \vspace{3pt}

\end{EPEreference}
\end{document}